# FeMnNiAlCr High Entropy Alloys with High-Efficiency Surface Oxide Solar Absorbers for Concentrating Solar Power Systems

*Xiaoxue Gao,[†] Edwin Jiang [†], Andrew Pike, Eldred Lee, Margaret Wu, Huan Wang, Sheppard Somers, Weiyang Li, Geoffroy Hautier, Ian Baker, and Jifeng Liu\**

[†]Authors contributed equally to this work
Thayer School of Engineering, Dartmouth College
15 Thayer Drive, Hanover, NH 03755, USA
*E-mail: Jifeng.Liu@dartmouth.edu

E. Lee is currently with Samsung Electronics, Inc.; M. Wu is currently with Frontiers Media SA; H. Wang is currently with Nankai University. S. Somers is current with Mastercard Data & Services

**Abstract**

High entropy alloys (HEAs) have attracted substantial interest in recent years. Thus far, most investigations have focused on their applications as structural materials rather than functional materials. In this paper, we show that FeMnNiAlCr HEAs can potentially be applied as *both* a structural and functional material for high-efficiency concentrated solar thermal power (CSP) systems working at >700°C. The HEA itself would be used in high-temperature tubing to carry working fluids, while its surface oxide would act as a high-efficiency solar thermal absorber. These HEAs have demonstrated yield strengths 2-3x greater than that of stainless steel at 700°C and a creep lifetime >800 h at 700°C under a typical CSP tubing mechanical load of 35 MPa. Their Mn-rich surface oxides maintain a high optical-to-thermal conversion efficiency of ~87% under 1000x solar concentration for 20 simulated day-night thermal cycles between 750°C and environmental temperature. These HEAs have also sustained immersion in unpurified bromide molten salts for 14 days at 750°C with <2% weight loss, in contrast to 70% weight loss from a 316 stainless steel reference. The simultaneous achievement of promising mechanical, optical, and thermochemical properties in this FeMnNiAlCr system opens the door to new applications of HEAs in solar energy harvesting.

# 1. Introduction

Solar energy is the most abundant renewable energy source available so far. Concentrating solar power (CSP) and solar photovoltaics are two major approaches to harvesting solar energy.[1] In a CSP system, sunlight is focused onto a receiver to heat up a working fluid, such as molten salts or supercritical $CO_2$ ($sCO_2$), which in turn drives a heat engine to produce electricity (**Figure 1**).[2] Compared to photovoltaics, CSP systems offer great advantages in cost-effective energy storage since the solar-heated working fluid can be stored and kept at a high temperature for >10 h. [3] This storage capability provides solar electricity even when the sun is not shining, resolving the intermittency issue of solar energy towards dispatchable solar electricity. [2] A higher operation temperature can drastically improve the energy conversion efficiency of CSP systems according to Carnot's Theorem. However, currently the operation temperature of commercial CSP systems with thermal storage is limited to <600ºC [4] due to challenges in high-temperature tubing materials and air-stable solar absorbers.

Materials for CSP tubing must have good creep resistance under high temperature operation, be able to both withstand the corrosion of working fluids and not oxidize excessively in air during operation. The requirements are somewhat similar to those for advanced supercritical power plants. The martensitic/ferritic alloys that are currently used in power plants are limited to operating temperatures ≤600°C. [5] Nickel-based superalloys, some titanium alloys, and, possibly, oxide-dispersion strengthened ferritic alloys [6, 7] can satisfy the strength and, at least for nickel-based alloys, the oxidation requirements at elevated temperatures. Unfortunately, these materials are too expensive except for use in specialized applications, [8] and corrosion at >700ºC by molten salts or $sCO_2$ still poses a significant challenge. Thus, besides the chemical and mechanical requirements listed above, we should add the requirement that the material is inexpensive for practical deployment in the CSP systems. Although ceramic tubing materials offer high-temperature stability at a relatively low cost, the intrinsic brittleness, lack of joinability/weldability, and the relatively low solar absorptance [9] limit their applications in CSP systems.

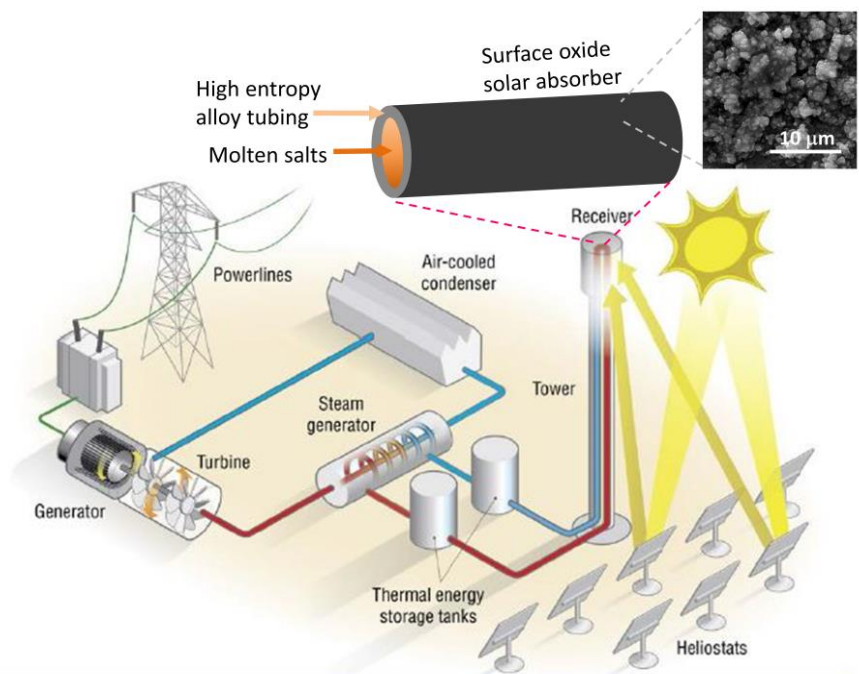

**Figure 1.** Schematic showing the application of FeMnNiAlCr HEAs to CSP receiver tubes. The HEA serves as the high temperature tubing material, while its native oxide on the outer surface (see the inset) serves as a high-efficiency solar absorber. The figure is adapted from the CSP schematics in the public domain from the U.S. Department of Energy following the corresponding policy of usage https://www.energy.gov/web-policies.

Another critical component of CSP systems is durable high-temperature solar absorbers, either as a surface layer or in a volumetric format. Ideally, such solar selective absorbers should absorb maximal solar radiation (wavelength $\lambda$=300-2500 nm) and convert it to heat with minimal thermal emittance losses in the mid-infrared (MIR) regime ($\lambda$ = 2.5-25 µm), thereby achieving a high optical-to-thermal energy conversion efficiency $\eta_{therm}$>90%. [10,11] The benchmark Pyromark 2500 solar paints currently used in CSP systems need multistep heat treatments for good adhesion, [12] and the optical-to-thermal conversion efficiency $\eta_{therm}$ is typically ~90% for 1000x solar concentration after annealing in air at 700°C for 480 h. [13,14] When operating at 750°C, the $\eta_{therm}$ decreases to ~88% within ~300 h due to phase transitions of iron oxide pigments and adhesion degradation. While our recent work has developed spinel oxide nanoparticle-pigmented high-temperature solar selective absorber coatings maintaining 94% efficiency at 750ºC, [15] it would be ideal if a high-efficiency solar absorber layer can be directly grown on the CSP tubing material for even better adhesion and more cost-effective fabrication.

In this paper, we investigate an innovative *synergistic* solution to the mechanical, optical, and thermochemical challenges of high-temperature CSP receivers using novel FeMnNiAlCr high entropy alloys (HEAs). Since the pioneering work of Cantor et al. [16] and Yeh et al. [17] on CoCrFeMnNi and CuCoNiCrAl$_{0.5}$Fe, respectively, there has been substantial and growing interest in HEAs, [18] which were originally defined as multicomponent alloys containing a minimum of five metallic elements with amounts in the range 5-35 at. %.[15] A later, somewhat arbitrary definition is that HEAs have a configurational entropy greater than 1.5R, where R is the gas constant. [19,20] Most studies have been focused on changing the substitutional atoms and determining how this affects the phases present and the mechanical properties as structural materials. In this paper, we show that FeMnNiAlCr HEAs [21,22,23,24] can potentially be applied synergistically as *both* structural and functional materials for high efficiency concentrated solar thermal power (CSP) systems working at ≥700ºC. The HEA itself would be used in high-temperature tubing to carry molten salts or sCO$_2$, while its Mn-rich surface oxide would act as a high-efficiency solar thermal absorber, as schematically shown in **Figure 1**. With Fe and Mn being the major components in these HEAs (adding up to ~70 at.% of the alloy), these materials are much more cost-effective than the Ni-based superalloys currently being investigated for high-temperature CSP systems, in which expensive Ni and Cr add up to ~70 at.% composition (**Table 1**). In fact, the total compositions of Ni and Cr in these HEAs are similar to those in stainless steels, while the yield strengths are 2-3x greater than that of 304 stainless steel from room temperature to 700 ºC, with a creep lifetime >800 h at 700ºC under a typical CSP tubing mechanical load of 35 MPa. Their Mn-rich native surface oxides have maintained a high optical-to-thermal conversion efficiency of $\eta_{therm}$~87% at 700°C under 1000x solar concentration ratio for >20 simulated day-night thermal cycles. Therefore, high-efficiency solar absorber can be achieved by surface oxidation of these HEAs instead of applying external solar coatings. In preliminary corrosion studies, two-phase Cr-modified HEAs have sustained immersion in unpurified bromide molten salts for 14 days at 750°C with <2% weight loss, in contrast to 70% weight loss from a reference 316 stainless steel sample. The simultaneous achievement of promising mechanical, optical, and thermochemical properties in this FeMnNiAlCr system enables new applications of HEAs in harvesting solar thermal energy.

**Table 1**. Compositions of single-phase C-doped HEAs and two-phase Cr-doped HEAs in comparison with those of stainless steel (SS) 304, 316 and Inconel 740H. The total atomic compositions of Ni and Cr in the Fe-Mn based HEAs are similar to those in the stainless steels and much less than Ni-based Inconel superalloys.

|  | Fe (at.%) | Mn (at.%) | Ni (at.%) | Cr (at.%) | Al (at.%) | Others (at.%) |
|---|---|---|---|---|---|---|
| **Single-phase C-doped HEA**[21,22] | 40.0 | 34.4 | 11.2 | 5.9 | 7.4 | C: 1.1 |
| **Two-phase Cr-modified HEA**[23,24] | 28.2 | 32.9 | 18.8 | 6 | 14.1 | N/A |
| **SS 304** | ~71 | 2 | 8 | 18 | N/A | Si: 0.75 |
| **SS 316** | 63-72 | Max 2 | 10-14 | 16-18 | N/A | Mo: 2-3 |
| **Inconel 740H** | N/A | N/A | 51 | 24.5 | 1.5 | Co: 20<br>Ti: 1.5<br>Nb:1.5 |

## 2. Mechanical Behavior of FeMnNiAlCr HEAs

In this section, we will summarize typical mechanical behavior of single-phase C-doped and two-phase Cr-modified FeMnNiAlCr HEAs, showing that their high-temperature mechanical performance is far superior to that of stainless steel 304 currently used in power plants.

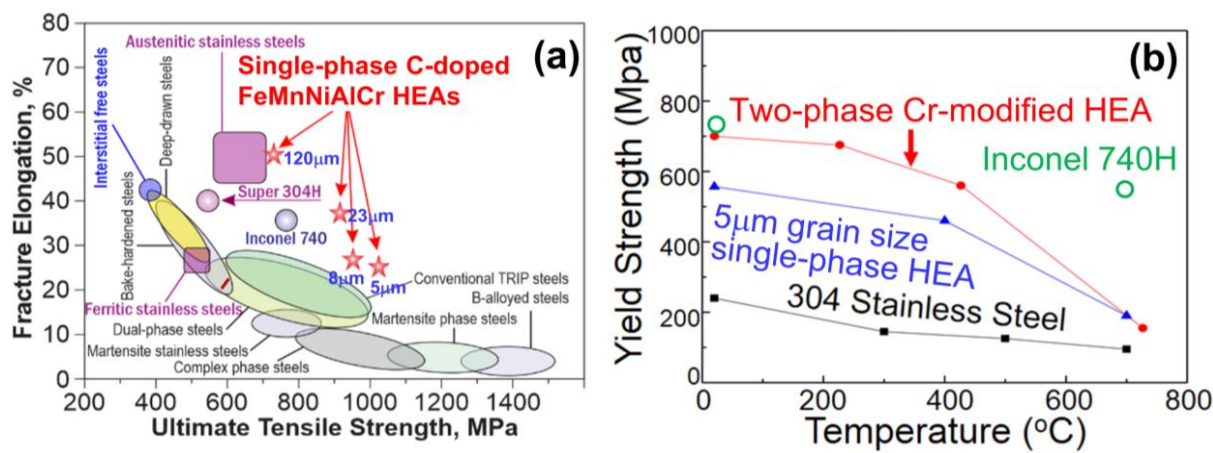

**Figure 2.** (a) Elongation vs. ultimate tensile strength at room temperature for a variety of steels, Ni-based Inconel 740 superalloy, and 1.1 at.% carbon-doped, single-phase $Fe_{40.4}Mn_{34.8}Ni_{11.3}Al_{7.5}Cr_6$ HEA recrystallized to different grain sizes (labelled in blue letters). (b) Yield strength vs. temperature for a two-phase Cr modified HEA and a recrystallized, single-phase C-doped $Fe_{40.4}Ni_{11.3}Mn_{34.8}Al_{7.5}Cr_6$ HEA (5-µm grain size) compared to that of 304 stainless steel and Inconel 740H. Data in the plots are collected from Refs. 21-25, 27.

We have shown in our previous work that the $Fe_{40.4}Mn_{34.8}Ni_{11.3}Al_{7.5}Cr_6$ (at. %) HEA can dissolve at least 1.1 at.% carbon interstitially while maintaining a random single-phase f.c.c. structure in the as-cast state, as confirmed by transmission electron microscopy (TEM) and atom probe tomography (APT). [21] This alloy will be referred to as single-phase C-doped HEA in this paper, and the atomic compositions with 1.1 at.% C is shown in Table 1. The carbon has an extraordinary effect on the mechanical properties, viz., it increases both the yield strength and the ductility: the room temperature yield strength increases linearly with carbon content and was found to be proportional to the increase in lattice strain.[22] Further, the carbon not only increases the work hardening rate, but the work hardening rate increases with increasing strain. These carbon-induced effects are related to the measured 50% reduction in stacking fault energy, and the onset of an unusual deformation mechanism at high strains, i.e. microband formation. [22] Recrystallization of this alloy at different elevated temperatures (>1000 °C) leads to different grain sizes and mechanical properties,[21, 26] i.e. the mechanical properties can be tailored. This is shown in **Figure 2a**, which also shows that the room-temperature tensile properties of the carbon-doped, FeMnNiAlCr HEAs compare very favorably to a wide variety of steels as well as Ni-based Inconel 740 superalloy, a state-of-the-art tubing material for CSP systems.[27] For example, $Fe_{40.4}Mn_{34.8}Ni_{11.3}Al_{7.5}Cr_6$ + 1.1 C HEAs with an average grain size of 23 µm exhibit a higher tensile strength (at a similar fracture elongation) than Ni-based Inconel 740 superalloys. Compared to stainless steel, the high temperature mechanical properties in the fine-grained (5 µm) single-phase HEAs are superior, see **Figure 2b** (the blue curve). The yield strength is 2-3x greater than that of 304 stainless steel from room temperature up to at least 700ºC.

In addition to the carbon-doped single-phase FeMnNiAlCr HEAs, we have found that two-phase alloy $Fe_{30}Mn_{35}Ni_{20}Al_{15}$ consisting of alternating (Fe, Mn)-rich f.c.c and (Ni, Al)-rich B2 (ordered b.c.c.) lamellae [28] have a good balance between strength and ductility at room temperature. [29, 30, 31] While this alloy suffers from environmental brittlement at room temperature due to attack by water vapor, adding Cr not only overcomes this environmental embrittlement issue but increases the overall elongation [23,24] For example, for the optimum addition of Cr, the room temperature elongation increases to ~18% at all strain rates with little change in yield strength compared to the Cr-free alloy. The lamellar spacing (500 nm) is unchanged by the Cr addition, with the most (~85%) of the Cr partitioning to the f.c.c. phase. The temperature dependence of the tensile properties of $Fe_{36}Mn_{33}Ni_{18}Al_{13}$ + 6 Cr two-phase HEA is also shown in **Figure. 2b**. Similar to $Fe_{40.4}Mn_{34.8}Ni_{11.3}Al_{7.5}Cr_6$ + 1.1 C, the yield strength is 2-3x greater than that of 304 stainless steel at all temperatures tested, while the elongation to failure is between 18-13% at all temperatures tested. It has also been shown that carbon additions (1.3 at.%) to $Fe_{36}Mn_{33}Ni_{18}Al_{13}$ can change the microstructure from lamellar-structured to a randomly-distributed two-phase microstructure in which the B2 phase changed to body-centered tetragonal. [32] Thermo-mechanical treatments of this alloy produced room-temperature yield strengths of 600 MPa with 25% elongation. Therefore, C-doping provides another handle to optimize the mechanical performance of the two-phase HEAs. Adding Ti to two-phase B2/f.c.c. FeNiMnAl HEAs has also been shown to lead to substantial improvements in yield strength, by refining the lamellae widths but at the expense of a reduction in ductility.[33] While the high-temperature yield strength of 250-300 MPa at 700°C is still lower than that of Inconel 740H (~560 MPa), it is already ~10x higher than typical mechanical loads in CSP tubing systems (25~35 MPa).[34]

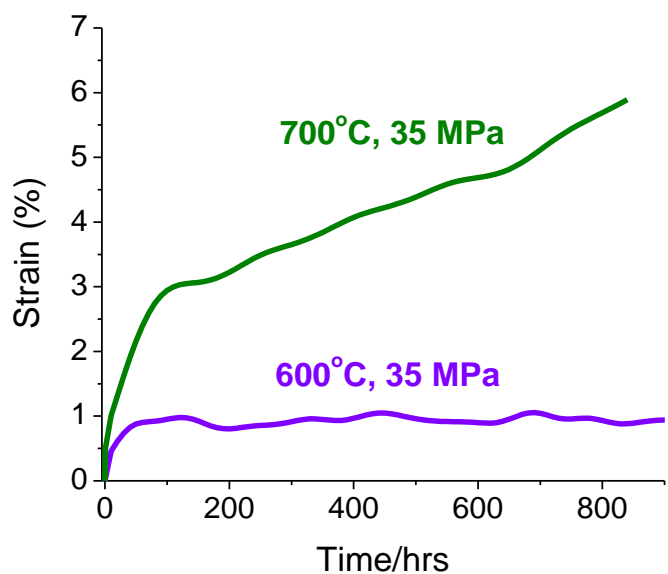

**Figure 3.** The creep curves for the two-phase Cr-modified FeMnNiAlCr HEAs in as-cast condition. The applied constant tensile stress is 35 MPa at 600 and 700°C.

We further conducted preliminary tensile creep studies at >600°C on the two-phase Cr-modified HEAs (as listed in **Table 1**). Choosing two-phase rather than single-phase HEAs for creep testing is mainly due to their superior surface oxide absorber and high-temperature corrosion resistance, as will be detailed in the following sections. Creep experiments were conducted at 600-700°C under a tensile stress of 35 MPa similar to the mechanical load of CSP tubing systems [34] utilizing a custom-built constant-stress creep apparatus designed following the approach of Garofalo, Richmond, and Domis [35]. As shown in **Figure 3**, the two-phase HEAs have sustained >800 h creep testing at 700°C under a stress of 35 MPa without failure. It is also promising to further optimize the creep resistance based on the alloying and microstructure engineering methods discused earlier. Further considering the dramatic cost reduction of these Fe-Mn based HEAs compared to Ni-based superalloys, these results indicate that FeMnNiAlCr HEAs are promising candidates for cost-effective high temperature tubing materials in high efficiency CSP systems.

## 3. Surface Oxide Solar Selective Absorbers on FeMnNiAlCr HEAs

### 3.1. Optical Properties of Native Surface Oxides on FeMnNiAlCr HEAs

Interestingly, the Fe-Mn-based HEAs investigated in this work also naturally offer an approach to efficient surface oxide solar absorbers. We have shown that both $MnO_2$ and $MnFe_2O_4$ nanoparticles have high solar absorptance in our previous work. [36,37,38] In fact, $MnO_2$ and $MnFe_2O_4$ have been used as black pigments since pre-Roman times in Ancient Italy, [39] known as "manganese black" in art history. This insight motivated us to explore the surface oxidation of FeMnNiAlCr HEAs for high solar absorptance without externally applied solar coatings.

Notably, the CSP system has its unique thermal cycles in which the entire system operates at high temperatures during the day, e.g. 700-750°C, and cools down towards the environmental temperature at night. The huge temperature difference of the thermal cycle could potentially induce severe spallation of the external solar absorber layer and severely reduce the overall optical-to-thermal conversion efficiency. For example, externally applied benchmark Pyromark 2500 ($\eta_{therm}$ =89.0-90.2%) high-temperature solar paints suffer from serious delamination and degradation after operating at high temperatures. [40] In order to observe the performance of our HEA native oxide solar absorber during the thermal cycles, we annealed the HEA samples at 800°C for 24 h in air to form the solar-absorbing surface oxide first, then test their optical performance after different durations of simulated "day-night" cycles. Each cycle includes 12 h at 750 °C in air and a 12 h natural cool down to room temperature.

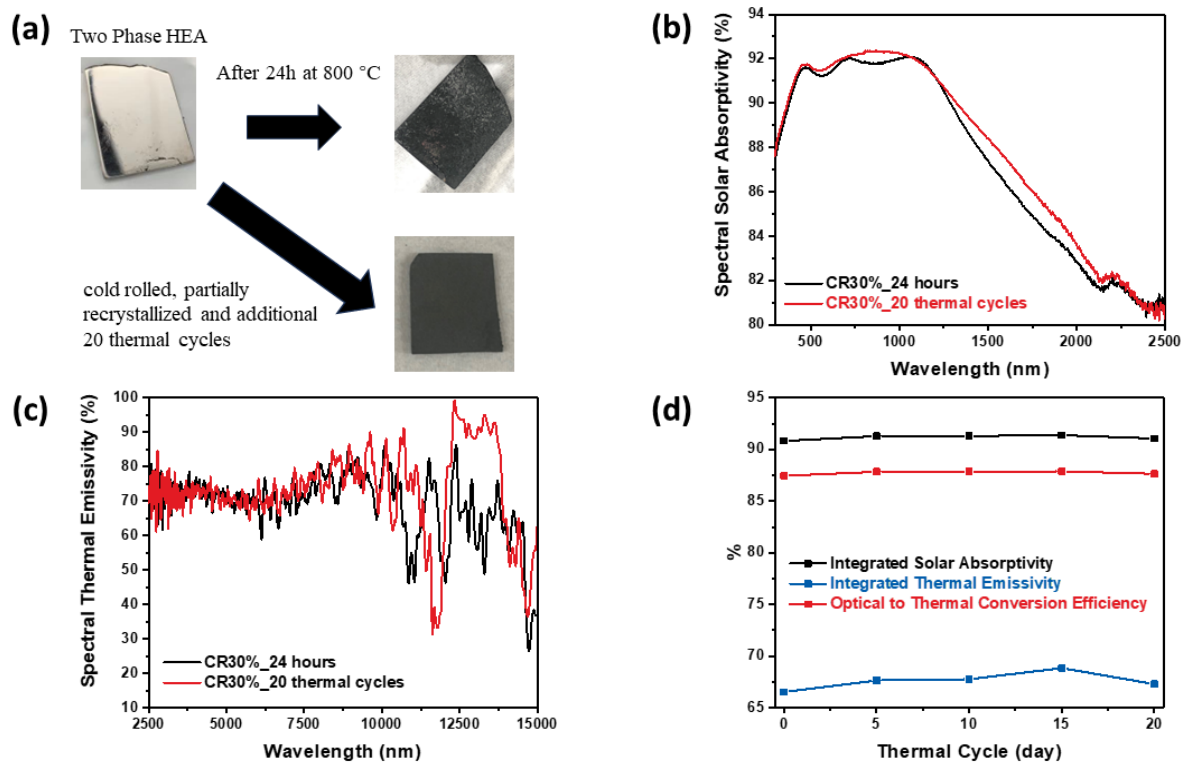

**Figure 4.** (a) optical photos of two-phase HEA before oxidation, after 24 hours annealing at 800 ºC and after cold rolled, partially recrystallized and additional 20 day-night thermal cycles. (b-d): optical characterizations of surface oxides on cold rolled and partially recrystallized two-phase HEAs. (b) spectral solar absorptivity; (c) spectral thermal emissivity; (d) spectrally integrated solar absorptivity, thermal emissivity, and overall optical-to-thermal conversion efficiency after 24 h at 800 ºC and an additional 20 simulated day-night thermal cycles between 750ºC and room temperature.

So far, the best surface oxide solar thermal efficiency and thermal stability were achieved by oxidizing cold-rolled two-phase Cr-modified HEAs with 30% thickness reduction followed by recrystallization annealing at 1000°C in Ar for 1 h. **Figure 4** demonstrates its optical performance in detail. **Figure 4a** shows the photos of as-cast two-phase HEA, after 24 hours annealing in air at 800 ºC, and cold worked and partially recrystallized two-phase HEA after 20 day-night thermal cycles in air. Without cold working and partially recrystallization, some oxide spallation could be observed from as-cast two-phase HEAs after only 24 hours annealing at 800 ºC. In contrast, with 30% cold working and annealing, the surface oxide layer is still uniform without any visible spalling even after 20 day-night thermal cycles. **Figure 4b** compares the spectral solar absorption of the cold-worked HEA after a 24-hour annealing and after 20 thermal cycles. Its spectral solar absorption even increases slightly after 20 thermal cycles. The solar spectral selectivity in **Figure 4b**, i.e. notably higher visible light absorptance than infrared absorptance, is mainly attributed to the mixed cubic and orthorhombic $Mn_2O_3$ phases, which have bandgaps of 0.55 eV and 0.79 eV, respectively, from Density Functional Theory (DFT) modeling with Perdew-Burke-Ernzerhof (PBE)+U. The band structure and optical properties of these two phases will be detailed separately in another paper. [41] **Figure 4c** demonstrates the HEAs' spectral thermal emissivity after 24 h and 20 day-night thermal cycles. The thermal emittance at λ<10 μm is almost unchanged. Since the peak black-body radiation wavelength at 700ºC is 2.98 μm, emittance changes at λ>10 μm have almost no impact on the overall spectral integrated thermal emittance at >700ºC.

To evaluate the optical-to-thermal conversion efficiency, we calculated the overall solar absorptance and IR thermal emittance from the data in **Figure 4b** and **Figure 4c**. The overall spectrally integrated solar absorptance can be calculated using the following equation: [10,11]

$$\alpha_{sol} = \frac{\int_{300\,nm}^{2500\,nm} I_{sol,\lambda} \alpha_\lambda d\lambda}{\int_{300\,nm}^{2500\,nm} I_{sol,\lambda} d\lambda} \tag{1a}$$

Where $I_{sol,\lambda}$ is the radiation intensity at a wavelength λ in a standard air mass (AM) 1.5 solar spectrum, and $\alpha_\lambda$ is the spectral absorptivity shown in **Figure 4b**. The overall thermal emittance at a temperature T is given by [10,11]

$$\varepsilon_{therm} = \frac{\int_{2000\,nm}^{20000\,nm} I_{black,\lambda} \varepsilon_\lambda d\lambda}{\int_{2000\,nm}^{20000\,nm} I_{black,\lambda} d\lambda} \tag{1b}$$

Where $I_{black,\lambda}$ is the black body radiation intensity and $\varepsilon_\lambda$ is the IR emissivity shown in **Figure 4c**. The optical-to-thermal conversion efficiency is given by [10,11]

$$\eta_{therm} = \alpha_{Sol} - (\sigma T^4 \varepsilon_{therm})/(CI_{sol,\text{int}}) \tag{1c}$$

Where $\sigma = 5.67\times10^{-8}$ W m$^{-2}$ K$^{-4}$ is the Stefan-Boltzmann constant, T is the temperature in Kelvin, C is the solar concentration ratio, and $I_{sol,int} = \int_{300\,nm}^{2500\,nm} I_{sol,\lambda} d\lambda = 1000$ W m$^{-2}$ is the spectrally integrated solar radiation intensity in a standard AM 1.5 solar spectrum.

**Figure 4d** shows the spectrally-integrated solar absorption, thermal emittance, and overall optical-to-thermal conversion efficiency versus the number of simulated day-night thermal cycles assuming a solar concentration ratio, C=1000 for solar power tower receivers operating at T=700°C. Interestingly, the HEA native oxide solar absorber maintains a high optical-to-thermal energy conversion efficiency of ~87% over the 20 simulated day-night thermal cycles without any degradation, which reveals the outstanding high-temperature cyclic thermal stability of this native solar absorber. Combined with the excellent high-temperature mechanical properties demonstrated in **Section 2**, these preliminary results indicate that FeMnNiAlCr HEAs show great promise for synergistically achieving the mechanical and optical performance required in high-efficiency CSP systems.

### 3.2. Microstructures of Surface Oxides on FeMnNiAlCr HEAs

Since the surface oxide on cold rolled and partially recrystallized two-phase HEA shows high solar absorptance and high optical-to-thermal conversion efficiency, we investigate the phases present and its microstructure. **Figure 5a** shows the complex surface topography of the native oxide on the two-phase cold-rolled and annealed HEA after a 24 h annealing at 800°C. These spontaneously-formed surface texture on the scale of microns is similar to that of solar cells,[42] which helps to reduce the solar reflectance and increase the solar absorptance in the HEA oxide layer. **Figure 5b** presents the surface oxide morphology after 20 day-and-night thermal cycles for comparison. There is little spallation of the native oxide observed on the surface which is consistent with its highly stable optical-to-thermal conversion efficiency upon high-temperature thermal cycling in **Figure 4d**.

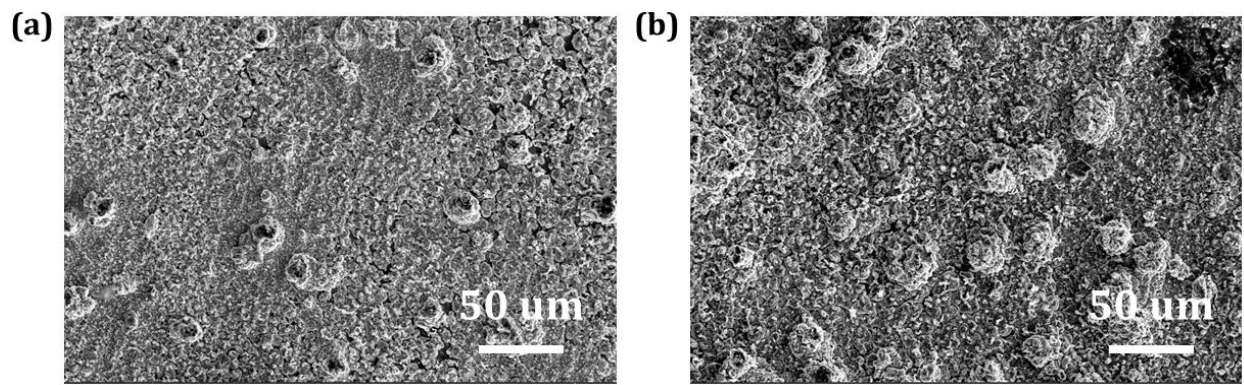

**Figure 5.** Morphology of surface oxides on cold-rolled and annealed two-phase HEAs: (a) after a 24 h anneal at 800 °C in air. (b) after additional 20 simulated day-night thermal cycles between 750ºC and room temperature.

X-ray diffraction (XRD) analysis was performed to identify the native oxide phases on the two-phase cold-rolled and annealed HEAs. **Figure 6** compares the XRD data of the HEA after a 24 h anneal at 800 °C and after additional 20 thermal cycles. The surface oxide is mainly composed of $Mn_2O_3$; f.c.c. and B2 phase peaks from the underlying alloy are also present for the sample oxidized for 24 h in air. After 20 thermal cycles, there is no obvious change of the oxide, which indicates its outstanding thermal stability during cyclic oxidation. The intensities of the f.c.c. and B2 peaks are greatly reduced after 20 thermal cycles due to the increased oxide thickness and, thus, reduced X-ray penetration into the underlying HEA.

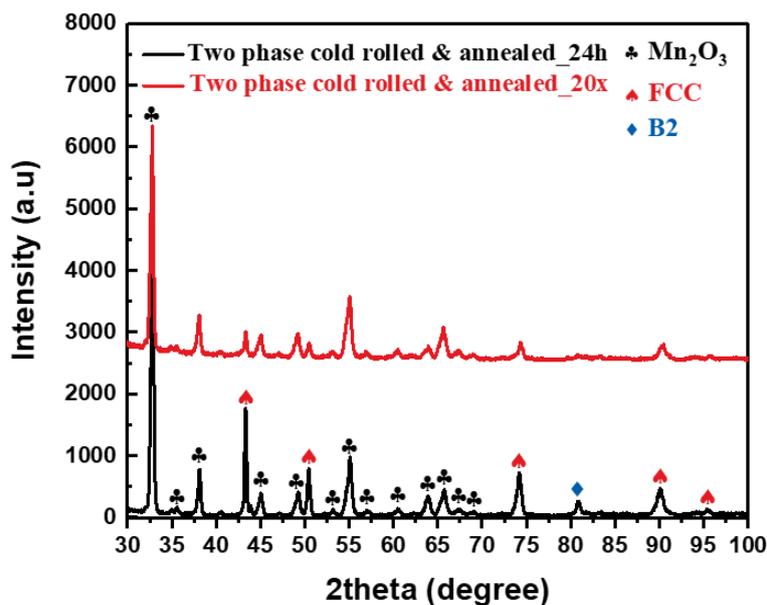

**Figure 6.** XRD data of two-phase cold rolled and annealed HEA after a 24 h anneal at 800 °C and after 20 thermal cycles.

To better understand the surface oxide layers, we use focused ion beam (FIB) milling to cut cross-sections on the oxidized two-phase HEAs to observe the oxide/alloy interface. A platinum protective layer was deposited on the oxidized HEA before FIB milling to avoid damage to the sample surface from the ion beam. Energy dispersive X-ray spectroscopy (EDS) elemental maps from a cross-section are shown in **Figure 7**. The surface oxide is clearly rich in Mn, which confirms our previous XRD data showing the presence of $Mn_2O_3$ diffraction peaks. Underneath this manganese oxide, there is a thin aluminum oxide layer along the alloy/oxide interface. The outermost layer of manganese oxide mainly contributes to solar absorption while the aluminum oxide along the alloy/oxide interface protects the underlying alloy from excessive oxidation. **Figure 8** shows elemental EDS maps from a cross-section of the two-phase cold-rolled and annealed HEA after 20 thermal cycles. The alloy/oxide interface is still robust and there are no obvious interfacial defects present. Even though the $Mn_2O_3$ layer thickness increased from ~4 μm to ~6 μm after 20 thermal cycles, the underlying protective alumina layer still remains continuous, which could be further engineered to improve the thermal stability of the native oxide in the future.

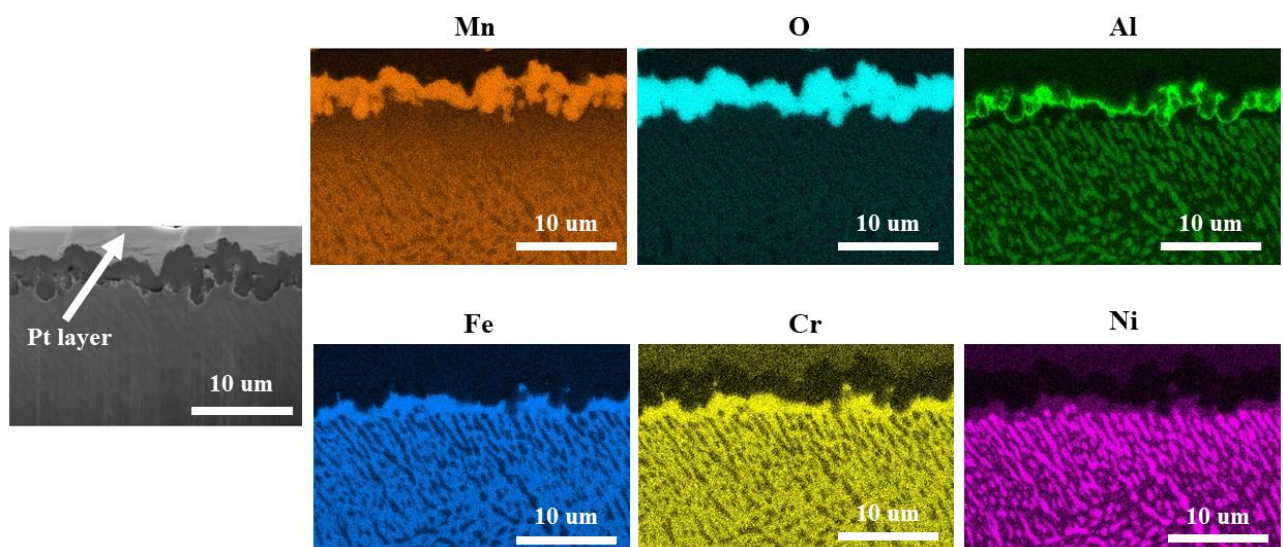

**Figure 7.** Secondary electron image and corresponding elemental EDS maps of a cross-section of the cold-rolled and annealed two-phase HEA sample after a 24 h anneal at 800 °C in air.

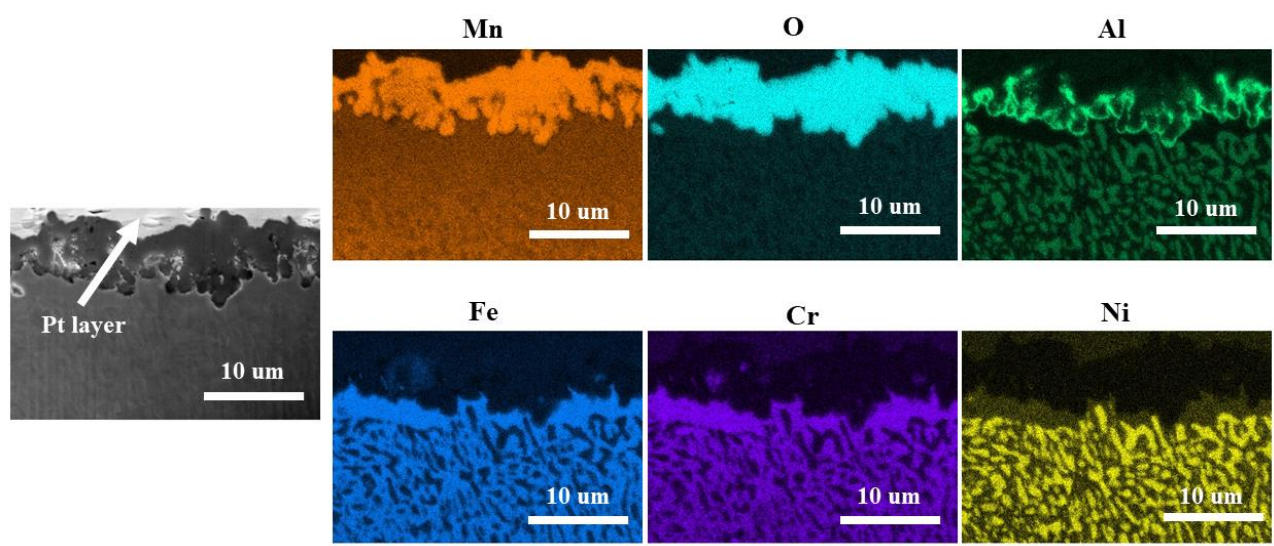

**Figure 8**. Secondary electron image and corresponding elemental EDS maps of a cross-section of the cold-rolled and annealed two-phase HEA sample after an additional 20 day-and-night thermal cycles.

**4. Preliminary High-Temperature Corrosion Studies of FeMnNiAlCr HEAs**

Another important aspect of high-temperature CSP tubing materials is the corrosion resistance to molten salts. Currently, chloride salts are the most promising candidates for CSP systems operating at >700°C due to their low cost. [34] However, these salts require elaborate purification before corrosion testing. [34] As an intermediate step, we tested the high-temperature corrosion resistance of HEAs in *unpurified* molten bromide salt with 67.1 mol.% KBr + 32.9 mol.% LiBr. These salts have a low melting point of 329ºC, similar to the chloride salts currently being investigated for high-temperature CSP systems. These salts are also chemically stable up to 1265ºC, and, thus, well suited for corrosion testing at 750ºC. We performed immersion corrosion measurement at 750°C for up to 14 days in the presence of air using the above bromide salt mixture for 316 stainless steel (commonly used for tubing materials due to good corrosion resistance), the single-phase C-doped FeMnNiAlCr HEA, and the two-phase Cr-modified FeMnNiAlCr. The samples have identical areas of 1.4 cm$^2$ to facilitate a direct comparison. The results are summarized in **Figure 9a**. While the reference 316 stainless steel sample lost 70% of its weight due to extensive corrosion, the single-phase HEA and the two-phase only lost 3.7% and <2% of their weight, respectively. Note that these HEAs are bare metals without any protection layers and the bromide salts have not been purified at all. We expect that the corrosion resistance will be further improved with $O_2$ and $H_2O$ removal [34] from the bromide molten salts (as required for their chloride counterparts). The superior corrosion resistance of Fe-Mn-based HEAs compared to stainless steel is especially intriguing since Mn is often unfavorable for high-temperature molten salt corrosion, yet this trend is reversed in FeMnNiAlCr HEAs, with the two-phase HEA performing better than the single-phase HEA. Therefore, there is plenty of space to further investigate the corrosion mechanisms in HEAs and engineer the HEA compositions and phases for optimal corrosion resistance at high temperatures.

To better understand the electrochemistry of the corrosion process, we have also used the potentiodynamic polarization method to measure the corrosion current and potential in sea salt solution (0.6 M NaCl aqueous solution). The electrochemical Tafel plots in **Figure 9b** show that the corrosion current of the single-phase HEA sample is less than half that of the 316 stainless steel sample. By comparison, the two-phase HEA corrosion current is even lower, at only one quarter that of the 316 stainless steel sample. These results also qualitatively agree with the high-temperature corrosion data in **Figure 9a**, showing that HEAs are much more corrosion-resistant than 316 stainless steel, and that the two-phase HEA performs even better than single-phase HEA. All these results show a great potential to further engineer and optimize the compositions and phases of Fe-Mn based HEAs towards corrosion-resistant high-temperature tubing materials for CSP systems.

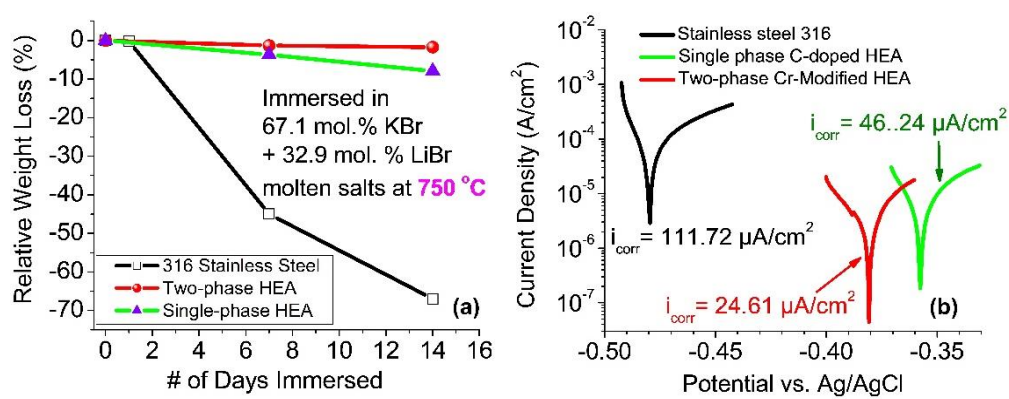

**Figure 9.** Comparison of the corrosion of 316 stainless steel, single-phase C-doped FeMnNiAlCr HEA and two-phase Cr-modified HEA: **(a)** in bromide molten salts at 750°C. All samples have an identical area of 1.4 cm$^2$; **(b)** Electrochemical Tafel plots measuring the corrosion current ($i_{corr}$) in sea salt solutions (0.6 M NaCl aqueous solution) using Ag/AgCl reference electrode.

## 5. Conclusions

In conclusion, we have shown that FeMnNiAlCr HEAs could potentially provide an innovative, cost-effective, and *synergistic* solution to the mechanical, optical, and thermochemical challenges of high-temperature CSP receivers, serving as both structural and functional materials. With Fe and Mn being the major components in these HEAs (adding up to ~70 at.% of the alloy), these materials are much more cost-effective than the Ni-based superalloys currently being investigated for high-temperature CSP systems, in which expensive Ni and Cr add up to ~70 at.% composition. The total compositions of Ni and Cr in these HEAs are similar to those in stainless steels, while the yield strengths are 2-3x greater than that of 304 stainless steels from room temperature to 700°C. The FeMnNiAlCr HEAs also endured more than 800 h creep testing at 700 ºC. Their Mn-rich native surface oxides have achieved a high optical-to-thermal conversion efficiency of $\eta_{therm}$~87% at 700°C under 1000x solar concentration ratio and could maintain this efficiency to > 20 simulated day-night thermal cycles. In preliminary corrosion studies, two-phase Cr-modified HEAs have sustained unpurified bromide molten salts for 14 days at 750°C with <2% weight loss, in contrast to 70% weight loss from a reference 316 stainless steel sample. The simultaneous achievement of promising mechanical, optical, and thermochemical properties in this FeMnNiAlCr system validates their potential applications in solar thermal energy harvesting.

**Experimental Section/Methods**

*Sample Preparation*:

*Bulk materials:* All FeNiMnAlCr HEAs materials were produced by arc-melting pieces of high-purity metals (> 99%) in a water-chilled copper crucible under an argon atmosphere. To compensate for the loss of manganese due to evaporation during the melting work, an additional 5 wt.% Mn was added to the mixtures. Each ingot was flipped and remelted three times to ensure homogeneous mixing. The recrystallized samples are prepared by machining blocks from the cast ingots to a thickness of ~6.5 mm and cold-rolled to various rolling reductions using a two-high, 4.2'' rolling mill with a small reduction (approximately 1%) per pass.

*Creep samples:* The creep specimens were milled into dog-bone shapes and then sliced to thin pieces by a high-speed diamond saw with gauge cross sections dimensions of 2.7 mm x 1.8 mm and a gauge length of 20 mm. Creep specimens were polished with fine silicon carbide paper up to 1200 grit, and then further polished with 5 μm and 1 μm alumina powder to a mirror finish.

*Sample for oxidation studies*: Ingots were cut into rectangular slices (around 1.5 cm x1.5 cm x1.5 mm) and grinded with 320 grit SiC paper. Before annealing in the box furnace, those pieces are ultrasonically cleaned with ethanol and dried by compressed air. Samples were annealed in the box furnace under laboratory air at 800 °C for 24 hours to form the solar absorbing layer first. After naturally cooling down to room temperature, they were taken out for characterizations. For the cyclic oxidation test, oxidized samples were annealed in the box furnace at 750 °C for 12 hours and naturally cooled down to room temperature for 12h in each cycle.

*Electron Microscopy*:

Scanning electron images (SEM, Helios 5 CX, Thermo Fisher Scientific, secondary electron (SE) mode) were employed to study the coating micromorphology. Focused ion beam (FIB) is used to prepare the cross-section on the sample. Energy dispersive X-ray spectroscopy (EDS, Ultim Max, Oxford Instruments) was carried out to detect the chemical composition.

*X-ray Diffraction:*

X-ray diffraction was conducted using Rigaku ultraX 18HB X-ray Diffractometer with Cu Kα radiation, λ=0.15406 nm at 40 kV and 300 mA. The continuous scan speed was 1.0 degrees/minute at a sampling width of 0.020°.

*Optical Characterization:* According to Kirchhoff's law, spectral emittance $\varepsilon(\lambda)$ at a given wavelength λ is equal to the spectral absorptance $\alpha(\lambda)$ at thermal equilibrium. For opaque substrate (such as metals) with zero optical transmittance, $\alpha(\lambda) = \varepsilon(\lambda) = 1 - R(\lambda)$, where $R(\lambda)$ is the spectral reflectance at wavelength λ [17]. The reflectance spectra in the wavelength range of λ = 0.3 ~ 2.5 μm for solar absorptance calculation were obtained by using a Jasco V-570 ultraviolet/visible/near infrared (UV/Vis/NIR) spectrometer equipped with a Jasco ISN-470 integrating sphere to capture both specular and diffuse reflection. As for the reflectance spectra in the mid-infrared (MIR) region (λ=2.5 ~20 μm) for thermal emittance measurement, a Jasco 4100 Fourier transformation IR (FTIR) spectrometer with a Pike IR integrating sphere was used.

*Corrosion Studies:*

The high-temperature corrosion resistance was measured using an immersion method in a molten salt consisting of 67.1 mol.% KBr + 32.9 mol.% LiBr in the presence of air. Before each test, the HEAs were washed with DI water and acetone followed by drying with nitrogen gas. The HEAs were weighted and placed in the molten salt at 750°C for different durations (up to 14 days). After the immersion tests, the HEAs were sonicated in DI water for 0.5 h, dried and weighted. Note that the average mass before and after immersion was based on three tests for each sample. The corrosion rate was evaluated using the following formula:

$$CR(\mu m/day) = 10^4 [WL/\rho AT]$$

where WL is the weight loss (grams), $\rho$ is density (g cm$^{-3}$) of the alloy, A is the total immersed area )cm$^2$), and T is the immersion duration (days).

The corrosion current and potential of the metal alloys was evaluated in 0.6 M NaCl aqueous solution with a potentiodynamic polarization method at room temperature. The HEAs, Pt mesh and Ag/AgCl were the working electrode, counterelectrode and reference electrode, respectively. The potentiodynamic polarization tests were performed on an electrochemical workstation (VMP3, Bio-Logic Science Instruments) at a scan rate of 0.2 mV s$^{-1}$ with a voltage range of -0.5 V to 0.5 V.

**Acknowledgements**

This work was supported by U.S. Department of Energy, Office of Science's Established Program to Stimulate Competitive Research (EPSCoR) under the Award Number DE-SC0021347.

Received: ((will be filled in by the editorial staff))
Revised: ((will be filled in by the editorial staff))
Published online: ((will be filled in by the editorial staff))

**ToC figure** ((Please choose one size: 55 mm broad × 50 mm high **or** 110 mm broad × 20 mm high. Please do not use any other dimensions))

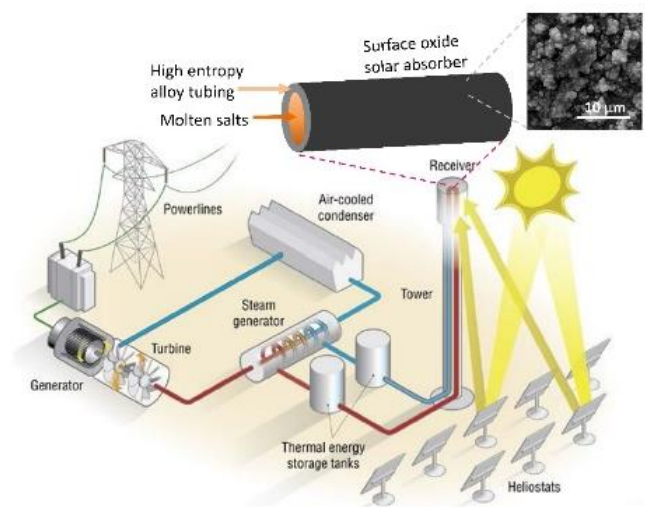